\documentclass[format=acmsmall, review=false, screen=true, anonymous=false]{acmart}

\usepackage{booktabs} 

\usepackage[ruled]{algorithm2e} 

\SetAlFnt{\small}
\SetAlCapFnt{\small}
\SetAlCapNameFnt{\small}
\SetAlCapHSkip{0pt}
\IncMargin{-\parindent}

\copyrightyear{2018}

\setcopyright{rightsretained} 
\acmJournal{PACMHCI}
\acmYear{2018} \acmVolume{2} \acmNumber{CSCW} \acmArticle{73} \acmMonth{11} \acmPrice{}\acmDOI{10.1145/3274342}

\received{April 2018}
\received[revised]{July 2018}
\received[accepted]{September 2018}

\begin{document}
\title[Addressing Bias \& Discrimination on Intimate Platforms]{Debiasing Desire: Addressing Bias \& Discrimination on Intimate Platforms}

\author{Jevan Hutson}
\affiliation{%
\institution{University of Washington School of Law}
	\city{Seattle}
	\state{WA}
  \country{USA}}
\email{jevanh@uw.edu}
\author{Jessie G. Taft}
\affiliation{%
  \institution{Cornell Tech}
  \city{NYC}
  \state{NY}
  \country{USA}}
  \email{jgt43@cornell.edu}
 \author{Solon Barocas}
 \affiliation{%
  \institution{Information Science, Cornell University}
  \city{Ithaca}
  \state{NY}
  \country{USA}}
    \email{sbarocas@cornell.edu}
\author{Karen Levy} 
 \affiliation{%
  \institution{Information Science, Cornell University}
  \city{Ithaca}
  \state{NY}
  \country{USA}}
    \email{karen.levy@cornell.edu}

\begin{abstract}
Designing technical systems to be resistant to bias and discrimination represents vital new terrain for researchers, policymakers, and the anti-discrimination project more broadly. We consider bias and discrimination in the context of popular online dating and hookup platforms in the United States, which we call \emph{intimate platforms}. Drawing on work in social-justice-oriented and Queer HCI, we review design features of popular intimate platforms and their potential role in exacerbating or mitigating interpersonal bias. We argue that focusing on platform design can reveal opportunities to reshape troubling patterns of intimate contact without overriding users' decisional autonomy. We identify and address the difficult ethical questions that nevertheless come along with such intervention, while urging the social computing community to engage more deeply with issues of bias, discrimination, and exclusion in the study and design of intimate platforms.
\end{abstract}

%
%
%
%


\maketitle

\renewcommand{\shortauthors}{J. Hutson et al.}

\section{Introduction}

Mobile dating and hookup platforms, which we call \emph{intimate platforms},\footnote{Previous scholarship uses various terms for different types of intimate platforms, often creating a dichotomy of platforms for finding love (dating apps) versus those for finding sex (hookup apps). Each has uses beyond these aims: friendship, community connection, entertainment. Importantly, histories of bias, discrimination, exclusion, subjugation, and power within intimate marketplaces transcend boundaries of platform categorization and call for a broader and more inclusive view of how intimacy is structured and practiced. Thus, we favor "intimate platforms" as an umbrella term because it allows the social computing community to more comprehensively grapple with the history, power, and consequences of intimate choice. Our analysis focuses on features that are consistent across platforms, though they may differ in user motivation.  } 
have become vital spaces where we meet and connect with intimate partners. Today 15\% of American adults report using dating sites and mobile dating applications \cite{pewdating}, with some research estimating that one-third of marriages start online \cite{cacioppo_marital_2013}. The impact on same-sex relationships is even more dramatic: one study found that in the late 2000s, over 60\% of same-sex couples met online \cite{rosenfeld_searching_2012}. Tinder and Grindr have tens of millions of users, with Tinder facilitating 20 billion connections since its launch \cite{match}.
 
Intimate platforms provide unparalleled access to prospective lovers and partners. Importantly, they present opportunities to create connections between formerly distant social groups. Rosenfeld and Thomas highlight how online dating "allows [people] to meet and form relationships with perfect strangers, that is, people with whom they had no previous social tie," displacing traditional patterns of intimate affiliation that involved close ties---such as families, neighbors, and mutual friends---and connections made locally in bars, neighborhoods, and workplaces \cite[p. 524]{rosenfeld_searching_2012}. This process of orienting individuals away from close ties and personal networks has the potential to disrupt existing patterns of assortative mating \cite{schwartz2013trends}, encouraging interaction between members of social groups that might have had little contact in the past. For example, some research suggests that online dating could increase rates of interracial marriage \cite{ortega2017strength}.  

While intimate platforms can provide new social opportunities, bias and discrimination may limit the degree to which such opportunities are realized in practice \cite{brown2018least}. Troubling research documents broad and pervasive inequities in the desire for, and appeal of, users of minority racial and ethnic backgrounds on intimate platforms \cite{rudder2014dataclysm,chowhu}. 

User profiles laced with phrases like "No blacks, sorry," "No Indians, no Asians, no Africans," or "Only here to talk to white boys," have given rise to critical research on and popular discourse about what many consider to be sexual racism \cite{callander_is_2015,bedi_sexual_2015,allen2015no,jones2016noasians}. Numerous platforms have come under fire for design practices that contribute to prejudicial modes of thought and action, such as race-based filters and search tools or race-based matching algorithms \cite{shadel2018dating,smothers_this_2016,strudwick_founders_nodate}. In light of these challenges, new intimate platforms, design practices, and marketing strategies are emerging that seek to intervene against user bias and discrimination, and more generally to promote social justice and equality \cite{locke2018gay,gremorequeerty}.

Designing technical systems to be resistant to bias and discrimination represents vital new terrain for researchers, policymakers, and the anti-discrimination project more broadly \cite{levy2017designing}. However, as a realm of private personhood, intervening in intimate affairs is controversial: romantic and sexual choices are understood as intensely personal, well outside the scope of preferences that platforms might feel justified in influencing \cite{whitney2017autonomy}. Such reluctance, we argue, rests on a fundamental misapprehension about both the nature of desire and the degree to which platforms can avoid exercising influence over our preferences in partners. The objects of our affections are shaped by our cultural environments, as well as the affordances of the platforms that facilitate intimate interactions. 

This article proceeds as follows. First, we consider critical and empirical work on the importance of intimacy, intimate platforms, and design. We then review three design strategies used by intimate platforms popular in the United States, drawing on theories of Queer and social justice-oriented HCI to explore their impacts on intimate discrimination. We argue that focusing on platform design can help alter troubling patterns of social interaction, without unduly interfering with individual intimate choices. We then consider the roles for designers and platforms, and the ethical issues that arise when designing intimate interventions. Ultimately, we call for the social computing research community to engage more deeply with issues of bias, discrimination, and exclusion in the study and design of intimate platforms. 

\section{Intimate Impact}
The intimate sphere is central to virtually all social life. It is the site of the "privileged institutions of social reproduction, the accumulation and transfer of capital, and self-development" \cite[p. 282]{berlant1998sex}. To that end, the intimate is inextricably tied to relations of power \cite{oswin2010governing}. Our intimate affiliations (and the structures that manage them) shape our selves, families, neighborhoods, communities, and nations. In the words of Berlant, "intimacy builds worlds" \cite{berlant1998}.

While individual intimate preferences are generally regarded as private matters that ought to be free from outside assessment and influence, \emph{systematic} patterns in such preferences---and the structures that promote and preserve these patterns---hold serious implications for social equality. As others have shown, the intimate sphere has historically been a crucial locus of state control, as well as a key determinant of social and economic welfare \cite{emens_intimate_nodate,green2016capital,bedi_sexual_2015}.

The intimate sphere has a particularly loaded social and legal history. The management of intimate affiliations and attachments played a critical role in colonial North America, helping to establish social distinctions and categories of identity that served as the basis for oppression and subjugation \cite{stoler2010carnal}. On numerous occasions, the U.S. government has enshrined particular forms of discrimination in intimate marketplaces, such as anti-miscegenation laws that prohibited interracial marriage and anti-sodomy laws that prohibited homosexual intercourse (see \cite{kitch2016anti,eskridge2008dishonorable} for review). Furthermore, the state has advanced a variety of policies that impact who we are (and are not) able to meet, interact, sleep, or fall in love with. For example, scholars highlight the U.S. government's role in redlining, residential segregation, school segregation, and the institutionalization of persons with mental and physical disabilities as "institutional examples of the state's shaping which intimate accidents can occur" \cite[p. 1380]{emens_intimate_nodate}, and argue that these restrictions have played a role in shaping intimate preferences and social norms in ways that impede certain relationships. These norms and relationships are then solidified over time, in a self-reinforcing loop that concretizes disadvantages for certain groups in intimate marketplaces and in the social world more broadly \cite{emens_intimate_nodate}. In this view, individuals' intimate affiliations are not the product of "pure" individual choice, but are instead shaped by accretions of state and social power.

Access to or exclusion from intimacy has a host of individual and structural outcomes. For individuals, intimacy is tied to a number of consequences for personal welfare, including (but not limited to) community attachment, sexual sociality, health, and well-being \cite{green2008social}. Extensive research has established a positive relationship between sexual activity and such outcomes as lifespan \cite{palmore_predictors_1982} and overall happiness \cite{blanchflower_money_2004}. Marriage is also associated with decreased mortality risk and improved health outcomes \cite{johnson_marital_2000}. Intimacy also has socioeconomic consequences: assortative mating has been shown to contribute to income inequality \cite{greenwood_technology_2015,bruze_male_2015}.

\section{Intimate Discrimination}

The intimate realm represents one of the only remaining domains in which individuals may feel entitled to express explicit preferences along lines of race and disability \cite{emens_intimate_nodate}. Even describing such preferences as biased or discriminatory can be challenging. As a matter of personal preference, sexual attraction might seem definitionally discriminatory: to have \emph{any} preference is to favor some people, and disfavor others, as potential partners. But describing desire as discriminatory is a way to capture more than the mere fact of sexual preference; it is a way to recognize and name intimate affinities that emerge from histories of subjugation and segregation. 

At the extreme, preference in potential partners might very well rest on racial animus and overt prejudice---a belief that those of a different race are unworthy of affection or respect. Or, an individual might limit intimate encounters to others that belong to her own race, on the belief that her race is categorically superior to others. Such preferences might be rightly described as sexual racism in the sense that they reflect generally racist attitudes as expressed in choice of romantic partners. Objecting to sexual racism is a way to object to racism more broadly, while recognizing the distinct harms that victims of racism suffer as a result of discrimination in the intimate sphere and the secondary effects that such experiences have on other crucial areas of their lives. But objections might also focus on the fact that sexual racism is likely to be particularly pernicious, given the significant deference that the choice of intimate partners tends to command in comparison to similarly biased decisions in other important domains \cite{whitney2017autonomy}.

Making claims of bias in sexual preferences is more difficult in the absence of overt prejudice \cite{dovidio2004nature}. If desire is the expression of some unconscious inner drive---a preference well outside control and beyond reason---then there seems little hope in calling out people for predilections that fall on racial lines. 
On this account, while we may choose our romantic partners, we do not choose whom we find attractive. Yet sexual preferences do not emerge from a psychological or cultural vacuum. The fact that ideals of beauty vary across time and place makes this rather obvious. Sexual preferences are historically and culturally contingent---as much the product of the social world in which one grows up as some biological imperative (see \cite{caplan2013cultural} for review). Cultural representations of love and sex inform our understanding of which intimate affiliations are acceptable and desirable, and they serve as the psychological material from which our preferences are fashioned \cite{green2016capital}. While culture does not instill desire in us, it so profoundly shapes the focus of our desires that preferences in romantic partners cannot be understood as simply a matter of individual and idiosyncratic choice. In this sense, sexual preferences might exhibit bias if they reflect prevailing representations of desirable partners---and if these representations demean, denigrate, or fetishize members of particular racial groups \cite{robinson2007structural}.

Resisting biased sexual preferences does not depend on the existence of some pure state of desire stripped of cultural influence. Instead, resistance simply requires recognizing that desire is malleable, and that such preferences can have a significant practical impact on the lives and livelihoods of marginalized populations \cite{green2016capital,bedi_sexual_2015}. Expanding people's sexual horizons does not mean overriding some "true" or innate preferences; it means intervening in the unavoidable and ongoing processes by which our preferences emerge in our interactions with our social and cultural environment. Not only do intimate platforms influence people's willingness to express and act on their sexual preferences, they actively shape those very preferences in the way they present potential partners. 

\section{Intimate Inequities Online}
While discrimination certainly occurred (and continues to occur) in offline spaces, intimate platforms provide easy-to-use features that allow users to act on such preferences. As platforms become centralized marketplaces for finding intimate partners, even small design decisions that impact matching behavior can have significant effects in the aggregate.

Extensive research documents disparities in racial preferences in online dating. As OKCupid founder and data scientist Christian Rudder stated of the matches formed on his site, "when you're looking at how two American strangers behave in a romantic context, race is the ultimate confounding factor" \cite{rudder2014dataclysm}. For example, white users of OKCupid are more likely to receive messages or have their messages responded to than their non-white peers, while Asian men and black women are least likely to receive messages or responses \cite{rudder2014we}. Heterosexual women of all races prefer white over nonwhite partners \cite{rudder2014we,tsunokai2014online}. White men and women of all ages are more likely to pursue dates with white rather than non-white partners \cite{lin2013mate} and are least likely to date outside their race \cite{robnett2011patterns}, while Asian and Latino men and women demonstrate comparable patterns of racial exclusion \cite{robnett2011patterns}. College students are more likely to exclude blacks, particularly black women, as possible dates \cite{bany2014gendered}. Black men and women are ten times more likely to message whites on an intimate platform than whites are to message blacks \cite{mendelsohn2014black}. The extent of self-segregation in online dating, however, is shown to peak at the first stage of contact: users are more likely to communicate across racial boundaries when reciprocating than when initiating romantic interest \cite{lewis18814}. Importantly, users who receive messages across racial boundaries engage in more new interracial exchanges than they would have otherwise \cite{lewis18814}. 

A wealth of empirical work documents racial discrimination among gay and bisexual users of online platforms in particular.\footnote{Sexual minority communities have been early and active adopters of online platforms for communication and intimate affiliation. Accordingly, a good deal of the literature on intimate platform user experiences has focused on these groups.} This research describes the many ways in which discrimination reinforces racial stereotypes, hierarchies, and social distance (e.g., \cite{callander2016not,paul2010internet}; see \cite{bedi_sexual_2015} for further review). 

In the context of intimate platforms, gay and bisexual men are more likely than their heterosexual counterparts to distinguish between potential sexual or romantic partners on the basis of race or perceived racial identity \cite{phua2003crossroads,plummer2007sexual}. In a 2012 study of racialized language in user profiles on Manhunt.net (a dating website for gay and bisexual men), researchers found that racialized language is leveraged in user profiles for a variety of purposes, including negative discrimination (e.g., "No Blacks."), positive discrimination or fetishization (e.g., "Black guys are sexy."), and marketing of the self to other users \cite{callander2012just}. Moreover, recent research demonstrates that discrimination between potential intimate partners on the basis of perceived racial identity is closely associated with generic racist attitudes (i.e., racism in non-intimate contexts) \cite{callander_is_2015}. 
These findings challenge the narrative of racial attraction as simply being a matter of innate personal preference by revealing a complex interaction between sexual politics and broader attitudes toward different racial and ethnic groups.

For minority racial groups, pervasive forms of rejection, discrimination and marginalization---whether on- or offline---can cultivate deep feelings of personal shame and lead subjects of discrimination to view themselves as less attractive or desirable \cite{callander_is_2015,caluya2006gay,han2007they}. A 2015 study found that 84\% of gay and bisexual men from minority racial backgrounds had experienced racism within gay communities, and 65\% of those respondents reported resultant stress \cite{han2015stress}. Another study found that Black gay and bisexual men's experiences with race-based discrimination in sexual interactions were significantly associated with sexual and relationship issues, including problems maintaining affection and finding a partner \cite{zamboni2007minority}. 

Even when people from minority groups are the object of sexual attraction, their desirability is often framed by stereotypical images \cite{phua2003crossroads}, often associated with particular racial bodies and identities. Research has documented perceptions of Asian men in white-dominated spaces as effeminate, submissive, and docile, forcing them to take on the "submissive" intimate role in hookup settings, and making them more likely to contract sexually transmitted infections, including HIV/AIDS \cite{han_qualitative_2008,chuang1999using,han2006think}. Similarly, experiences of social discrimination were found to predict likelihood of engaging in risky sexual behavior, and resulting HIV transmission, among gay Latino and Black men \cite{ayala2012modeling}. These findings draw a direct connection between sexual racism and public health outcomes.

\section{Justice, Design, and Intimate Platforms}
The fields of human-computer interaction and social computing have long histories of considering large-scale techno-social issues of power and justice, and actively promoting pro-social design solutions. HCI theorists and researchers have worked to promote positive online community-building \cite{kraut2012building}, environmental sustainability \cite{disalvo2010mapping}, civic engagement \cite{harding2015hci,korn2015creating}, bystander intervention \cite{difranzo2018upstanding} and social justice generally \cite{dombrowski_social_2016} in a variety of contexts and using a variety of technologies. 

Many newer HCI theories promote a more critical and activist approach to design, encouraging debate and questioning of historical scripts and norms by both designers and users in such a way that makes these theories ideal for tackling issues of intimate discrimination. Historically, much HCI work has had the goal of making user interfaces as easy to use as possible---supporting "natural" behaviors, without questioning why they are considered to be so. According to one seminal work, "HCI practice reflects and reproduces existing relations between groups of people, whatever they happen to be" \cite{light_hci_2011}. Queer, feminist, and postcolonial theories of HCI, however, encourage "troubling" and reimagining historical relations as mechanisms for seeing outside the design status quo \cite[p.23]{light_hci_2011, philip2012postcolonial}. These theories encourage the use of design to question what is natural, challenge the boundaries created by the histories of heteronormativity, and inspire design that helps to expand users' intimate horizons \cite{kannabiran2012designing,marshall2016user}. 

The importance of critical engagement with intimate platform design has been noted in other fields. Legal theorists have understood intimate platforms by analogy to physical social institutions, like bars, schools, and houses of worship \cite{levy2017designing,robinson2007structural}. These physical spaces operate as "architectures of intimacy" that determine both which potential romantic partners one is likely to meet, and attitudes about the type of potential partners worthy of one's attention \cite{emens_intimate_nodate}. Political theorists also argue that intimacy is a matter of justice, as access to meaningful intimate affiliation can be critical to accessing primary social goods such as wealth and self-respect \cite{bedi_sexual_2015}. They describe how the racial steering and screening tools available on intimate platforms reify extant racial stereotypes and hierarchies, impacting the likelihood that users express and act on racial preferences, while contributing to the social pressure on people of color to conform to racialized sexual stereotypes \cite{robinson2007structural,bedi_sexual_2015}. They note that discrimination-enabling design practices would  run into legal trouble in other domains, such as housing and employment, and even suggest "regulating web site design decisions that produce, exacerbate, or facilitate racial preferences" \cite[p. 2794]{robinson2007structural}.

Empirical research confirms these worries. For example, sexual minority men who use intimate platforms more frequently viewed multiculturalism less positively and sexual racism as more acceptable \cite{callander_is_2015}. Researchers attribute these findings to the use of "simplified racial labels" in profile design and in search and filter tools, contending that these design choices "encourage the belief that [simplified racial labels] are useful, natural or appropriate for defining individuals and sexual (dis)interest" \cite[p. 1998]{callander_is_2015}. Other research describes how intimate values are "scripted" into platforms by founders and designers, constructing an ideal "desiring user" \cite{hardy_constructing_2017}. The experiences and identities of culturally and geographically diverse users can clash with the platform's idea of who a user should be, causing the script to break down.

Considering intimate platforms as architectures of intimacy empowers designers to intervene in issues of bias and discrimination at the structural level: changing the underlying affordances \cite{norman2013design} of platforms without directly prescribing or proscribing users' intimate decisions. Users who harbor intimate biases, whether conscious or not, may well continue to make intimate decisions informed by these biases. Individual decisions, and their outcomes, can still be discriminatory, fetishizing, or otherwise marginalizing regardless of the structural design solutions in place. However, these approaches to HCI encourage designers to abandon the pretense of neutrality in favor of promoting overall social good. Design interventions of this type do not remove from users the ability to make intimate decisions on the basis of protected characteristics, but instead frame users' choices and expand intimate possibilities in a way that works toward equity and justice for all users.

\section{Design Features for Mitigating Bias}
How might intimate platforms be designed to mitigate bias and discrimination in user behavior? In this section, we highlight current design features on intimate platforms that may impact the degree to which preferences on the basis of race (or other protected characteristics) affect whether and how users encounter one another. We identify three prominent design features common on intimate platforms: \emph{search, sort, and filter tools}; \emph{matching by algorithm}; and \emph{community policies and messaging}.\footnote{This review focuses on platforms that have received scholarly attention or media coverage related to bias and discrimination. This is not to suggest that discrimination on these platforms is especially pronounced or that other platforms do not warrant critical assessment. Rather, the dynamics that we identify in these examples are likely to be common across platforms.} This list is by no means exhaustive; we selected these features for discussion based on the attention they have received in the political theory literature \cite{robinson2007structural,bedi_sexual_2015}, on their relation to Queer and feminist HCI theories (e.g. \cite{light_hci_2011}), and on our own previous research about the implications of platform design features on user-to-user bias and discrimination in other contexts (e.g., in market exchange, job-seeking, or ride-sharing \cite{levy2017designing}) .

\subsection{Search, Sort, and Filter Tools} 
When creating an account on an intimate platform, users are often encouraged (and in some cases required) to categorize themselves according to a number of characteristics.\footnote{For example, of the 25 top grossing dating and hookup apps on the iOS App Store in the United States (in March 2018)\cite{appannie}, 19 requested that users input their own race or ethnicity; 11 collected users' preferred race/ethnicity in a potential partner. 17 allowed users to search, sort, or filter others by the race or ethnicity.} Users are then allowed to search for, sort by, or filter out potential partners based on these self-described characteristics. Search and filter functions allow users to specify what they do or do not wish to see in potential mates, and return only users that meet those requirements. Sorting functions allow users to specify which characteristics are more or less desirable, and return an ordered list of potential partners. 

Search, sort, and filter tools rely on users placing themselves, and the partners they seek, into platform-defined categories. A platform's provision of certain categories (and the concomitant exclusion of others) for searching, sorting, and filtering legitimizes such categories as socially reasonable bases for including or excluding potential partners \cite{bowker2000sorting}. On one hand, these labels can be useful for letting users (including those in marginalized groups) self-identify according to particular characteristics and find one another. The converse implication is that these design features reduce the diversity of the field of potential matches \cite{finkel2012online}. They allow users to explicitly or implicitly exclude, demote, or fetishize others on the basis of race or other protected characteristics. These features cause the users excluded from a search to become invisible: they are screened out of the "dating pool" before they are even recognized as potential participants. 

Many of the sortable categories provided on intimate platforms, such as age, gender, and sexual orientation, likely strike us as reasonable grounds for sorting potential romantic or sexual matches. However, several intimate platforms also allow sorting by race, ethnicity, and HIV status. 

These design choices reify---and tacitly validate---extant stereotypes related to race, ethnicity, and other categories \cite{bedi_sexual_2015,robinson2007structural}. They map onto historical notions of psychological and physical group difference, and promote these categories as both natural characterizations of other users as well as appropriate axes for determining romantic or sexual (dis)interest. Screening tools based on protected characteristics undermine the potential of intimate platforms to bridge social distance \cite{rosenfeld_searching_2012,ortega2017strength}, as they allow users of different social or economic backgrounds to be made invisible. Harms may be magnified when platforms make these features available only to users who have paid extra for "premium" service, as doing so expands the agency of the socioeconomically privileged, while undermining the capacity of filtering tools as a means for users of lower socioeconomic status (often racial and ethnic minorities and persons with disabilities) to do the same.

By including design features that permit screening on the basis of these protected characteristics, platforms provide individuals with a form of control over their selection of potential partners, but do so in a way that naturalizes discriminatory preferences. The inclusion of filters that, for example, exclude users of certain races implicitly presents such preferences as normal and acceptable. It need not be so. Rather, platforms could seize an opportunity to challenge users' pre-existing notions through thoughtful design.

Instead of prioritizing perceived control to exercise preferences on the basis of protected characteristics, platforms could introduce friction or "seams" into the sorting and filtering process. These design philosophies emphasize the importance of making the technologies and decisions underlying an interface visible, adding opportunities for users to pause, reflect on their behavior, and build their understanding of a design product's internal logic \cite{chalmers2003seamful}. Making the process of screening based on protected characteristics more difficult or "seamful" could encourage users to consider potential partners as individuals undefined by race, ethnicity, or ability, rather than as nuisances to be screened out. 

Here, Queer HCI's emphasis on exploring beyond preconceived notions of what is desirable can directly inform platform design. Rather than allowing users to search for what they think they want, platforms could remove filtering features along these axes entirely or provide results that intentionally introduce diversity into the results displayed to a user. Diversity metrics are already common features in search and recommendation engines, and could also be incorporated into intimate platforms (e.g. \cite{vargas2011rank}).

Alternatively, platforms could replace standard sorting features with new ones that help users categorize others along new axes, or into categories that are less burdened by extant biases and stereotypes. These categorization issues have long been found in the the world of online pornography, whose search terms, content tags, and other organizing principles reflect stereotypical categories of race, sex, age, and ability. The sorting and filtering features that allow users to find the pornography of their choice are highly specific, and may encourage users to think of the videos they search for in racialized or otherwise discriminatory terms. Just as potential partners on intimate platforms shape intimate preferences \cite{robinson2007structural}, consumption of sexually explicit materials has been shown to influence sexual preferences, including along racialized lines (see for review \cite{dines2010pornland}). 

In response to these concerns, adult-film stars Stoya and Kayden Kross launched the porn site TrenchcoatX.com \cite{trenchcoatx}, which notably removes racial tags from its content and racial categories from its search mechanism \cite{trenchcoat}. Instead, users are required to search for individual performers or specific acts. This design decision introduces friction by making it more difficult for users to find content that they desire. But, in so doing, it compels the user to separate a performer from stereotypical search terms, and to individuate performers who might otherwise be known or seen through the lens of categories like "black MILF" or "Latina teen." This encourages users to look beyond existing fetishized categories in their pornography consumption, which may in turn encourage them to look beyond these categories in their intimate lives more broadly.

Other intimate platforms have included features that encourage users to de-categorize and re-categorize themselves and each other according to characteristics other than race, ethnicity, and ability. Many platforms include filters based on other non-demographic personal characteristics and interests, like education, political views, relationship history, and preferences around smoking and drinking. Platforms might also categorize potential partners along new axes unassociated with protected characteristics. The Japan-based gay hookup app 9Monsters groups every user into one of nine categories of fictional "monster" according to a process that includes both the user's own type preference and the community's perception of them \cite{9monsters}. While 9Monsters may still sort users into categories along established lines like body type or weight, it's possible that this re-categorization may also help users look past other forms of difference, such as race, ethnicity, and ability. If the platform allows users to cultivate common ground based on newfound affinity groups rather than existing categories, new forms of connectedness and identification may result \cite{finkel2012online}. In cases where searching, filtering, and sorting mechanisms are based on stereotypical categories, these new modes of identification and categorization may function to unburden historical relationships of bias in the intimate realm and encourage connection across existing boundaries.

\subsection{Matching by Algorithm}
The widespread use of algorithms to match users of intimate platforms with one another can introduce more subtle forms of bias and discrimination. Often, users are asked to complete profiles or surveys soliciting details about their own characteristics and their preferences for those characteristics in others. These often include ratings (weights) of the importance of those characteristics in users' intimate preferences. Users are then presented with potential mates who are recommended as "good matches" by the platform, presumably based on the preferences and weights specified.

While algorithm-driven approaches promise to simplify the matching process, they also bring many of the same risks that accompany the sorting and filtering features previously discussed. To the extent that matching algorithms rely on users' stated preferences for protected characteristics, these mechanisms can reify group differences and naturalize historically fraught decision criteria for selecting romantic partners.

However, algorithms' opaque operation and potentially biased inputs can further magnify these harms. When the screening process is automated, users may be unable to determine precisely how their matches were selected, or why others were deemed incompatible and thus made invisible. Users may assume that their stated preferences had some impact on their outcomes, but the use of aggregate user preferences to make match predictions can make the logic behind intimate matches difficult to understand \cite{finkel2012online}.

The structural assumptions and potential harms of algorithmic matching are made salient in the case of the dating app CoffeeMeetsBagel. Like many such apps, it allows users to specify both their own race and their preferred race(s) in a partner. In 2016, a series of well-publicized incidents \cite{notopoulos_dating_nodate} demonstrated that the app's matching algorithm tended to show users potential partners of their own race---even when they stated that they had no preference as to the race of matches. A spokesperson for the app clarified that "data shows even though users may say they have no preference, they still (subconsciously or otherwise) prefer folks who match their own ethnicity. [The algorithm] does not compute "no ethnic preference" as wanting a diverse preference" \cite{notopoulos_dating_nodate}.

This response reveals two underlying assumptions about automating intimate interactions. First, it assumes that, despite an explicitly stated assertion of a lack of racial preference, a user's \emph{inferred} desires should instead dictate with whom they are matched. Specifically, it suggests that because other similar users preferred matches of their own race, this presumed preference should supersede the user's expressed choice to be matched with users of different races \cite{tene2017taming}. In the case of CoffeeMeetsBagel, the platform has implemented an algorithm that provides users with a conservative interpretation of what they might seek. The user's stated choice is to be open to exploration, and instead of embracing this preference and offering diverse or creative suggestions that expand the user's field of potential matches, it instead intentionally narrows that field based on inferred preferences. Second, it assumes that a platform and its algorithm should strive to reproduce the preferences and choices of existing users. There is no indication to users of what data are used to form recommendations, or how such data were processed or analyzed. Platforms like these define a "good" future match by using the definition of a "good" past match, without considering how those past matches came to be.

There may be no easy way for platforms to establish a natural baseline of user compatibility against which to measure the effectiveness of their matching algorithms. Some intimate platforms have tried to establish such a baseline experimentally: OKCupid, for instance, ran a series of experiments in which users were (unwittingly) presented with "bad" matches (according to the site's algorithms) but told that they were highly compatible \cite{rudder2014we}. The site found that the power of suggestion had a real effect on user behavior: users who were predicted to be incompatible had more successful interactions when they were told by the platform that they were compatible. Though OKCupid's experiment faced public criticism (e.g. \cite{wood2018criticism})---primarily because users were not informed and did not consent to it---its results suggest that intentional experimentation with matching algorithm design can not only help users explore their preferences, but may be necessary in order to determine the efficacy of these tools and to discover unexpected good matches. As in many other contexts, algorithmic reliance on users' previous matches fail to account for the biases in those data and stand poised to reproduce those biases.

Instead, intimate platforms could take this opportunity to explore the new interactions, and new data, that exploration could generate. In the words of pioneering Queer HCI theorist Ann Light, this strategy would let "(other) values and lifestyles surface---not the ones already in use, but ones that might come to be if allowed enough space to emerge" \cite[p. 433]{light_hci_2011}. Here, preferences are disrupted through exploration: instead of relying on assumptions about subconscious preferences, on un-imaginative algorithms, or on questionable historical patterns, intimate platforms should instead encourage accidents and exploration with the goal of actively counteracting bias. 

Research suggests that people's intimate preferences are somewhat fluid, and are shaped both by the options presented to them and through encounters with things they don't expect \footnote{Platforms taking exploratory approaches should, however, assess whether doing so might create risks for marginalized users. On Tinder, for example, people who were surprised to find themselves matched with transgender users sometimes subjected those users to harassment and hate speech \cite{levy2017designing}. Platforms should holistically consider the affordances of their platforms when designing "surprising" experiences.} \cite{gerlach2017predictive,green2016capital}. Facilitating new or surprising experiences can encourage people to explore beyond their conscious (or even subconscious) preferences \cite{rudder2014we,finkel2012online}. Instead of using algorithms that choose the "safest" possible outcome, matches could be calculated with significantly more enthusiasm for diversity. This is not only true for people who have no preference as to the race of their potential partner. Matching algorithms could also subtly encourage people with explicit preferences to look beyond what they thought they wanted. 

\subsection{Community Policies and Messaging}
Some platforms attempt to address issues of overt discrimination or other antisocial behavior by including rules against it in their community policies.\footnote{Of the 25 top-grossing dating and hookup apps on the iOS app store in the United States (in March 2018)\cite{appannie}, 5 platforms included explicit anti-discrimination community policies or profile guidelines. }
 A common approach is the inclusion of community policies or profile guidelines that specifically warn against inappropriate behavior or promote respect and openness. Others use behavior agreements, profile badges, or shareable media articles to establish positive behavior norms and promote meaningful inquiry into intimate preferences (for review of the use of such mechanisms to mitigate bias on platforms, see \cite{levy2017designing}). 

While discrimination based on race or ethnicity is a top concern for many of these platforms, disability is also a salient concern. Individuals afflicted with HIV/AIDS, for example, face significant obstacles in social and intimate marketplaces. Popular gay dating and hookup applications are working with HIV/AIDS organizations to develop strategies to reduce the stigma associated with HIV \cite{sanfranciscoaidsfoundation_building_nodate}. For example, DaddyHunt, a location-based real-time dating and hookup application for sexual minority men, informs its users of the stigma and alienation experienced by users afflicted by HIV, and offers users the opportunity to indicate whether they are "open to dating someone of any [HIV] status" \cite{daddyhunt}. It then gives users the option to sign a "Live Stigma-Free" pledge, and if they choose to do so, adds a visible indicator of this pledge to users' profiles with the text "[user] lives Stigma-Free," as described in \cite{levy2017designing}.

By encouraging education and voluntary affirmation of this pledge, DaddyHunt prompts users to reflect on their own preferences. Instead of simply asking users for their serostatus or that of their preferred partners, the pledge signals to users that understanding and openness are important platform norms. It then visibly marks the profiles of users who promise to uphold these norms, allowing users to signal to each other both their own intimate preferences and their broader social attitudes. Because the pledge is entirely voluntary, DaddyHunt can encourage a more inclusive approach to HIV without coercion. 

Beyond seropositive stigma, platforms could encourage users to reflect upon the consequences of expressing a preference for a particular race or ethnicity (whether in their profile or in their match preferences). One approach is to explicitly prohibit certain user behaviors: gay dating app Hornet, for example, bars its users from including any language referring to race or racial preferences in their profiles or bios \cite{hornet}.

Platforms could also play a role in informing users of the extent of disparities faced by racial and ethnic minorities in online dating marketplaces---potentially including the harms to dignity that accompany both fetishizing and marginalizing behavior toward people of certain races, or the socioeconomic harms of assortative mating. Grindr's newly launched media arm "Into More" opens possibilities for intimate platforms to proactively engage these issues \cite{intomore}. Upon successfully logging onto the Grindr platform, users are greeted with a preview of one of the daily "Into More" stories. These stories range from current events to fashion and public health, and often engage specific issues relating to bias and other problematic stereotypes that involve Grindr's community of users. For example, the Into More article "14 Messages Trans People Want You To Stop Sending On Dating Apps" \cite{clements14messages} explores the biases and problematic behaviors that Grindr users exhibit toward trans users. It explains how describing trans people as a personal fetish, which many see as a positive expression of sexual preference, is inherently dehumanizing, and how negative preferences against intimate affiliations with trans people (as well as "disabled folks, fat folks, [and] femmes" \cite{clements14messages}) are the result of broader systems of subjugation and oppression. 

Design features that encourage users to explore their own preferences and biases challenge the presumption that intimate platforms have no role to play in shaping the social norms that create bias and discrimination. We contend that encouraging individual users to critically reflect on their preferences may support larger social norms that call biases into question. By encouraging inquiry and education via community guidelines and messaging, platforms may help their users look beyond existing categories or stereotypes and toward more diverse possibilities.
\section{Roles for Designers and Platforms}
Because so much intimate interaction begins in online spaces, the design features of intimate platforms---whether search, sort, and filter functions, matching algorithms, or community guidelines---necessarily shape the intimate possibilities that are available to each user. Platforms may emphasize their roles as facilitators of individual preference, attempting to remain "neutral" in their actions and allowing users to enact their own desires. However, platforms have no choice but to decide whether to include or exclude certain informational categories or present certain types of users to others as potential partners. No design choice is neutral: even attempts to cede as much control as possible to users does not relieve platforms of their powerful roles in structuring mediated social interactions \cite{gillespie2010politics}. Claims of neutrality from platforms ignore the inevitability of their role in shaping interpersonal interactions that can lead to systemic disadvantage. They also ignore well-established theoretical concepts, such as affordances, seamfulness, and friction, that describe the behavior-changing power of design \cite{norman2013design,chalmers2003seamful}.

If we understand the position of platforms to be necessarily non-neutral, asking that designers rethink their products to proactively minimize discriminatory outcomes for all users begins to seem like a less radical position. Design and social computing researchers have already called for more activist approaches to design, encouraging questioning of norms by both designers and users, and prompting debate on how designers imbue their creations with their own values \cite{branham_co-creating_2014,disalvo_making_2014}. Other work promotes efforts that recognize unjust practices, and hold those who enable such practices responsible \cite{dombrowski_social_2016}. These theories encourage designers to create tools that promote cooperation, equality, and good citizenship among users of all types, and can be extended to intimate platforms to advance pro-social outcomes.

Designers of intimate platforms have several other advantages that put them in an ideal position to intervene in discrimination. First, they have the scale required to address these issues. Users' intimate decisions on such platforms compound: the cumulative effect of millions of individual decisions can have profound impacts on the nature of intimate affiliation in society more broadly. Therefore, it is reasonable to expect the architects of these platforms to ensure that those preferences are examined and considered such that they do not unduly map onto historical patterns of discrimination.  

Second, design has advantages over other methods of intervention, like legal approaches, in mitigating intimate discrimination. In contexts covered by United States civil rights laws (like housing and employment), many of the categories we discuss here---including race, gender, and HIV status---constitute protected classes. While some offline establishments that facilitate relationship formation may be subject to anti-discrimination laws (e.g., guaranteeing accessibility and preventing discrimination on the basis of protected characteristics), these often do not apply to online platforms. Legal scholars have begun to argue for the application of such laws to platforms like Uber and Airbnb \cite{leong2016new}, but their application to intimate platforms has not been established; it is unlikely that federal civil rights laws would be construed to protect directly against users' discriminatory behavior on intimate platforms, and we do not intend to argue that they should do so.

We note, however, that there is some possibility that broader \emph{state} civil rights statutes might be construed to protect against discrimination on intimate platforms. For example, the dating website eHarmony was sued under California's Unruh Civil Rights Act, which grants Californians the right to "full and equal accommodations [... and] services in all business establishments"; the plaintiffs claimed that eHarmony violated the law by only allowing users to search for opposite-sex partners on their site. The case was resolved via a settlement agreement in which eHarmony agreed to provide an alternative same-sex matching service \cite{eharmony}. There are few useful lessons to draw from this case, as it was settled rather than adjudicated. eHarmony's apparent violation was also sufficiently extreme that it does not provide obvious guidance to more complicated situations: eHarmony's lack of a mechanism to search for same-sex partners amounted to an effective prohibition of gay users from the platform, whereas the design elements we discuss in this paper might allow marginalized populations to make use of the platform, but have a diminished experience there. Although we consider this case a fairly weak signal about the applicability of state civil rights law to the forms of intimate discrimination we discuss here,\footnote{Another legal framework often invoked regarding online discrimination is Section 230 of the Communications Decency Act, which grants platforms immunity against users' discriminatory (or otherwise illegal) conduct. However, platforms \emph{may} be liable if their designs structure users' interactions in ways that "help[] to develop" such conduct. In the most influential case interpreting Section 230, \emph{Fair Housing Council of San Fernando Valley v. Roommates.com} (521 F.3d 1157 (9th Cir. 2008)), the 9th Circuit Court held that a roommate matching platform was not immune because its drop-down menus \emph{required} users to state preferences for gender and sexual orientation, and its search and filter functions then steered users toward those elicited preferences. This conduct was considered to violate the Fair Housing Act, which forbids discrimination in housing. Section 230 is legally inapposite in the context of intimate platforms, as there is no federal law that prohibits discrimination in intimate partner selection. But it may be the case that platforms' concerns about maintaining "neutrality," derived from this jurisprudence, generally influence their willingness to intervene purposively on users' choices.} it raises important questions as to whether design features like search and filter tools---which enable users to systemically exclude others on the basis of protected characteristics---deprive certain users of full and equal services on intimate platforms. 

Ideally, the architects of these platforms should not be solely responsible for problems of discrimination; these challenges also require the engagement of policymakers, independent designers, researchers, and activists, with input from the communities affected by issues of bias and discrimination. However, in the absence of guidance from legal or governmental authorities, platforms themselves may be in the best position to take the lead in mitigating harms to marginalized groups.

\section{The Limits \& Politics of Intimate Intervention}
Despite the potential for pro-social intervention through design, there are a number of ethical issues that must be considered when designing these interventions. By no means do we claim to resolve these tensions, nor to stake out a specific normative prescription for platforms' decision-making. Rather, we suggest that anti-bias interventions on intimate platforms must, at minimum, contend explicitly with the issues we highlight here.

\subsection{Respect for autonomy and agency}
In suggesting that intimate platforms view design as a mechanism to combat bias and discrimination, we are encouraging them to consciously and purposefully intervene in the intimate decision-making practices of their users. While all design choices necessarily impact the horizon of possibilities available to users \cite{tong2016online}, such interventions might be viewed as an unjustified attempt to shape users' attitudes, beliefs, and ultimate decisions in matters of sex and love. Even platforms that do not force decisions on users nevertheless raise questions of autonomy; the power to influence users' choices is enough to court controversy \cite{rudder2014we}.

While such reactions might give platforms pause---perhaps appropriately so---they cannot escape responsibility for the influence they already wield. Users cede an enormous amount of agency to intimate platforms, with the expectation that the user's ideal outcome (e.g., a successful match) is the same as the platform's. In reality, exactly what those outcomes are and how platforms should go about realizing them may be ambiguous. As the case of CoffeeMeetsBagel \cite{notopoulos_dating_nodate} reveals, when algorithms find that daters' stated preferences are different from the characteristics of the people with whom they choose to interact, which should take precedence? Either choice can be understood as betraying users' intentions and interests.

Resistance to anti-bias interventions that rest on a strict notion of autonomy assumes that platforms know and understand how to satisfy users' true preferences \cite{whitney2017autonomy}, yet recent research has found that accurately predicting compatibility between two people using even sophisticated machine learning techniques is challenging \cite{joel2017romantic}. In contrast, limiting the ease with which users can act on certain preferences in the interest of reducing bias rests on the belief that preferences are never completely determined in advance. No user is a completely independent agent, with perfect self-knowledge to act in ways that accord with their pre-established preferences. Rather, preferences take shape and evolve in the process of navigating these platforms \cite{green2016capital,finkel2012online}.

Even if a platform takes the position that users arrive with pre-established preferences over which the platform itself has no influence, it may feel justified in discouraging users from \emph{expressing} certain preferences. For example, it may adopt policies that prohibit users from stating explicitly that they are uninterested in others of a particular race (e.g. "No blacks, sorry") or only willing to engage with those of their same race (e.g., "Only here to talk to white boys"). Platforms may not expect to \emph{change} users' preferences by limiting their ability to express these preferences, but platforms may do so nonetheless in the interest of protecting the dignity and sense of self-worth of other users. When confronting such messages, users who are (or are not) members of the identified racial group are likely to feel demeaned and unwelcome. 

In other words, a platform may view users' preferences as their own, while still rejecting the idea that users from minority groups must face denigrating messages as a condition of navigating the platform. It may enforce this policy by simply communicating that such behavior is unacceptable, or it may take affirmative steps to police users' expressions, perhaps limiting their ability to employ certain terms in their profiles or relying on similar keyword-based techniques to prompt manual review by human moderators \cite{levy2017designing}. Of course, limiting users' ability to express certain preferences may not prevent users from maintaining or acting on these preferences. In fact, it may do little to change how marginalized populations ultimately fare on the platform, as they may be no more successful in finding interested or receptive partners. But the cumulative effect of reducing the incidence of encountering such exclusionary messages may be substantial if such changes make marginalized users more likely to use the platform.

However platforms approach these issues, they should be transparent with users about their policy and design choices---as well as their rationales for these choices. Doing so can mitigate concerns about manipulation, provide users with greater agency to select platforms that work for them, and can provide additional incentive for other platforms to be more reflective about their own practices and goals.

\subsection{Histories of aggression and harmful intervention}
While social and political histories of subjugation are grounds for intervention to counter contemporary discrimination, they also ought to give us pause about the consequences of such intervention. Active intervention in the realm of intimate affiliation carries with it a litany of historical baggage that warrants substantial caution. Through anti-miscegenation laws, anti-sodomy laws, conversion therapy regimes, or forced sterilizations of persons with disabilities, state actors and policies have deprived many of the requisite agency and dignity to participate in intimate opportunities \cite{emens_intimate_nodate,kitch2016anti}. Although platform design presents an opportunity to engage intimate bias and discrimination in ways that are less bound up in issues of state power and violence, we cannot ignore the pitfalls of history; as such, we must consider how structural interventions through design map onto historical efforts to intervene in intimate marketplaces. 

Simultaneously, we ought to consider how some of the design features we describe here that can serve as vehicles for bias may also be of tremendous utility to historically marginalized groups. For instance, users whose identities are highly stigmatized might find search and filter functions essential for achieving intimacy, ensuring safety, and building community; offline and online sites specifically for gay communities have been incredibly important for these purposes \cite{taylor_social_2017,van2014breaking}. Decisions to remove, alter, or preserve design features should consider and balance the potential harms of inclusion against this utility.

\subsection{Which categories are appropriate for intervention?}
While it may strike us as normatively acceptable to encourage intimate platform users to be open to more diverse potential partners, we might find some categories more palatable for such intervention than others. For example, it might seem inappropriate to suggest that a Jewish user seeking other Jewish people "expand her horizons" past those preferences, which might be based on a number of religious and cultural considerations. Similarly, a platform suggesting that a gay user "consider" dating someone of a different gender would likely strike us as problematic. Intimate platforms can be very useful for minorities looking to meet others who share their background and values. Instead of drawing a bright line on what should or should not be acceptable categories to consider, we suggest that designers should take the needs of marginalized or historically oppressed populations into account when considering how intimate platform features are used. Careful consideration of the outcomes of the exercise of intimate preferences may reveal that some of these groups are at greater risk for harm than others, and that platform features should be implemented accordingly. 

\section{Conclusion}
As intimate platforms grow, so does their influence on individual and structural outcomes within and beyond the intimate sphere. Intimate platforms already intervene in and mediate our private lives and decisions, and many are beginning to work toward counteracting the bias and discrimination that often manifests in our most personal interactions. Recognizing the serious challenges and opportunities presented by design practices among popular intimate platforms, we advocate for the social computing research community to engage intimate platforms as critical structures for intervening in and reflecting on issues of bias, discrimination, and exclusion. Exploring structural rather than individual interventions allows us to move beyond debates of individual preference versus discrimination, and toward conversations that are concerned with how these preferences and practices arise and are sustained. Future research in the social computing community could work towards building design mechanisms that mitigate extant forms of discrimination, developing theories and practices of design that unsettle entrenched preferences, and unpacking complex legal and policy questions about the design, operation, and role of intimate platforms. 

\section{Acknowledgements}
We thank Anna Lauren Hoffmann, Amanda Levendowski, Tarleton Gillespie, Christine Geeng, and the anonymous reviewers for valuable conversations and feedback.

\bibliographystyle{ACM-Reference-Format}
\bibliography{dbd}

\end{document}